\documentstyle[12pt,graphicx]{article}

\title{Nucleation phenomena\\
       in a nonuniform atomic fluid\\
       in the electrical field}

\author{O.\,V.\,Derzhko$^{1,2}$,
        V.\,M.\,Myhal$^2$\\
        $^1$Institute for Condensed Matter Physics\\
        1~Svientsitskii Str., Lviv UA--79011, Ukraine\\
        $^2$Chair of Theoretical Physics, Ivan Franko National University of Lviv\\
        12~Drahomanov Str., Lviv UA--79005, Ukraine}

\begin{document}

\maketitle

\begin{abstract}
The density functional approach is used to study the gas--to--liquid
and liquid--to--gas nucleation phenomena in a fluid of two--level atoms in an
external electrical field. The influence of the field on the surface
tension and nucleation and cavitation barriers is discussed.



\end{abstract}

\vspace{5mm}

In recent years a significant progress has been made
in the molecular theory of the gas--to--liquid and liquid--to--gas phase transitions.
An important contribution in this field is that of D. W. Oxtoby with coworkers \cite{001}
who showed how the experimentally measured quantity,
the nucleation rate,
can be calculated starting from the interparticle interactions
with the help of the density functional approach.
The elaborated method seems to be valuable near the spinodal curve
since it predicts a vanishing of the nucleation barriers
(as it should be expected)
in contrast to the well known classical nucleation theory
which yields a finite value for the nucleation barriers.
The results for the nucleation rate obtained within the framework of those theories
differ near the binodal curve as well,
however,
a prediction of the classical nucleation theory in that case seems to be preferable.
Both theories can be viewed as microscopic ones
which are based on the density functional approach study
of the planar interface (classical nucleation theory)
or the spherical interface (nonclassical nucleation theory).

In the present paper we are following the influence of an external electrical field
on the gas--to--liquid and liquid--to--gas nucleation phenomena.
First we examine the changes in the interparticle interactions due to the field
adopting for this purpose a simple model of two--level atoms.
After deriving the equation of state we make a guess
about the form of the grand thermodynamic potential density functional,
which takes into account the external electrical field.
Finally, we performed standard density functional theory calculations
to obtain the surface tension and nucleation barriers
and discuss the influence of the external electrical field on these quantities.

To study the effective (long--range) interparticle interaction
in the presence of the external electrical field
we introduce two--level atoms
with the excitation energy $E_1-E_0$
and the value of the dipole transition moment $\vert{\bf {p}}\vert$.
Using the quantum mechanical perturbation theory
one immediately finds the long--range interparticle interaction between two atoms
being at the distance $R_{12}$
\begin{eqnarray}
\label{001}
- \left( \frac{\alpha_{12}^2}{2}
+\frac{\gamma_1^2+\gamma_2^2}{4} \left(1+\frac{3\alpha_{12}^2}{2}\right)
-\gamma_1\gamma_2\alpha_{12}\right) (E_1-E_0) +\ldots\;.
\end{eqnarray}
Here
\begin{eqnarray}
\label{002}
\alpha_{ij}(E_1-E_0)=\frac{\vert{\bf {p}}\vert^2}{R_{ij}^3}\Phi_{ij},
\;\;\;\;\;
\gamma_i(E_1-E_0)=-2\vert{\bf {p}}\vert\vert{\bf {E}}\vert\chi_i,
\end{eqnarray}
and
$\Phi_{ij}$ and $\chi_i$ are the known in the theory of dipole systems
orientational functions depending on the orientations of dipole transition moments.
The first term in (\ref{001}) is the usual van der Waals interaction,
all other terms depend on the strength of the electrical field $\vert{\bf{E}}\vert$.
We do not discuss a change in the short--range interaction due to the field
taking it into account by means of the ``atomic radius'' $\sigma$.

We calculate further the equation of state
repeating a text--book derivation of the second virial coefficient
for the case when the long--range interaction between two hard--sphere--like particles
is given by Eq. (\ref{001}).
The only peculiarity
is the additional average over the orientations of dipole transition moments
which we perform by means of the cumulant expansion.
Although the calculations are rather tedious
the final result for the virial state equation is simple \cite{002}
\begin{eqnarray}
\label{003}
&&
\frac{p}{kT} =\rho+\rho^2B_2 +\ldots,
\\ \nonumber
&&
B_2 = 4v -2\pi\sigma^3\int_{2\sigma}^{\infty}{\mbox{d}}\rho\rho^2
\left(\exp\left(\frac{\aleph^2}{3\tau} \left(1+2\aleph^2{\cal{E}}^2\right)
\frac{1}{\rho^6}\right)-1 \right),
\end{eqnarray}
where
$v=\frac{4}{3}\pi\sigma^3$
and
$\tau=\frac{kT}{E_1-E_0}$
and
${\cal{E}}=\frac{\vert{\bf{E}}\vert\sigma^3}{\vert{\bf{p}}\vert}$
are the dimensionless temperature
and the dimensionless strength of the electrical field,
respectively.
Within the adopted approximations
the particle is characterized only by the dimensionless parameter
$\aleph=\frac{\vert{\bf{p}}\vert^2}{\sigma^3\left(E_1-E_0\right)}$
In further computation we put for simplicity $\aleph=1$.
As can be seen from Eq. (\ref{003}) the only result of the external electrical field
is the increasing of the van der Waals constant by $1+2\aleph^2{\cal{E}}^2$.

It is easy to construct the grand thermodynamic potential density functional
\cite{003}
which yields for the uniform fluid of small density
the virial state equation (\ref{003}).
Thus, we shall examine the atomic fluid in the external electrical field
on the basis of the following grand thermodynamic potential density functional
\begin{eqnarray}
\label{004}
\Omega [\rho({\bf R})]
&=&
kT\int{\mbox{d}}{\bf R}_1\rho({\bf R}_1)
\left( \ln\left(\Lambda^3\rho({\bf R}_1)\right)
+\frac{-1+6v\rho({\bf R}_1)-4v^2\rho^2({\bf R}_1)}
{\left(1-v\rho({\bf R}_1)\right)^2}
\right)
\nonumber\\
&-&
\frac{6\sigma^3a}{\pi}
\int_{\vert{\bf R}_1-{\bf R}_2\vert\ge 2\sigma}
{\mbox{d}}{\bf R}_1{\mbox{d}}{\bf R}_2
\frac{\rho({\bf R}_1) \rho({\bf R}_2)}
{\vert{\bf R}_1-{\bf R}_2\vert^6}
-\mu\int{\mbox{d}}{\bf R}_1\rho({\bf R}_1),
\end{eqnarray}
where
$\Lambda$ is the thermal wavelength,
and the constant
\begin{eqnarray}
\label{005}
a=\frac{1}{48} v (E_1-E_0) \aleph^2 (1+2\aleph^2{\cal {E}}^2),
\end{eqnarray}

\begin{figure}[h]
\vspace{8mm}
\centerline{\includegraphics[width=10cm]{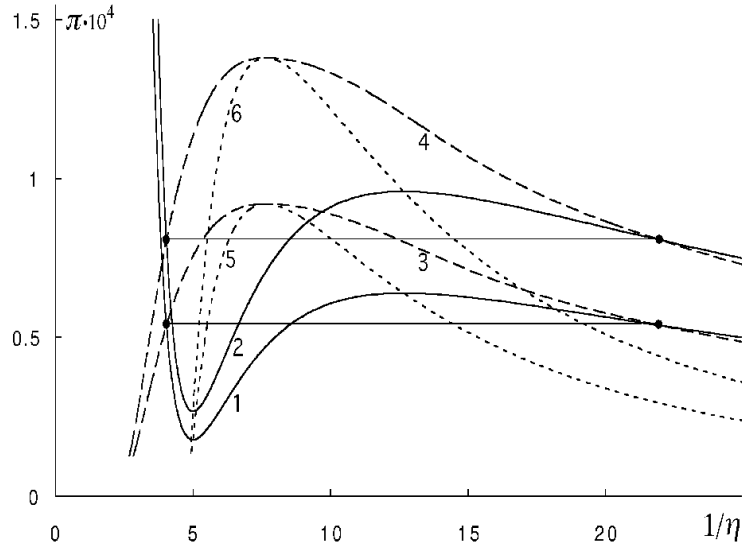} }
\caption{\small Isotherms
                $\pi=\pi\left(\frac{1}{\eta}\right)$
                ($\pi=\frac{pv}{E_1-E_0}$ and $\eta=\rho v$
                are the dimensionless pressure and density, respectively)
                at $\tau=0.9\tau_c({\cal{E}})$ (solid curves),
                binodal curves (long--dashed curves)
                and spinodal curves (short--dashed curves)
                in the absence of electrical field (1, 3, 5)
                and in the presence of electrical field
                ${\cal {E}}=0.5$ (2, 4, 6).}
\vspace{8mm}
\end{figure}

\noindent
which controls the contribution of the long--range interaction to
$\Omega [\rho({\bf R})]$,
depends on the strength of the field.
The grand thermodynamic potential density functional (\ref{004}) takes into account
the short--range interaction within the (local) Carnahan--Starling approximation
and
the long--range interaction within the (nonlocal) mean--field approximation.
We also consider a simplified version of (\ref{004})
treating the long--range contribution within the local gradient approximation
which is expected to yield a reasonable result
for the temperatures only slightly lower than the critical temperature.

We proceed by applying to (\ref{004})
the standard machinery of the density functional theory \cite{003}
(see also \cite{004,005})
seeking for a solution of the equation for density
$\frac{\delta\Omega[\rho(\bf r)]}{\delta\rho(\bf r)}=0$
and evaluating the grand thermodynamic potential $\Omega$.
We examine the uniform fluid
establishing the gas--liquid phase diagram
and the nonuniform fluid with planar and spherical interface
calculating the surface tension and the nucleation barriers
for the saturated vapour and the tensile liquid.
Our findings are collected in Figures~1--4.

\begin{figure}[h]
\vspace{6mm}
\centerline{\includegraphics[width=12cm]{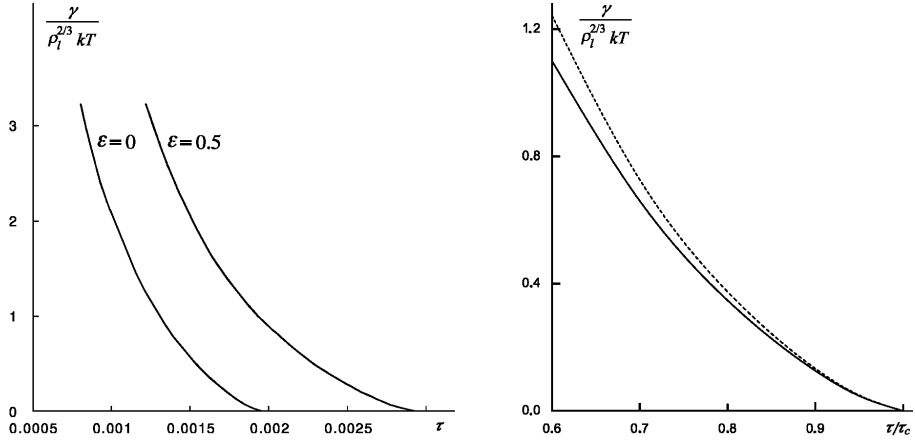}}
\caption{\small Temperature dependence of the surface tension $\gamma$.
                In the left panel
                the dependence $\gamma(\tau)$ for ${\cal {E}}=0$ and ${\cal {E}}=0.5$
                is shown.
                In the right panel the result of the nonlocal treatment (solid curve)
                is compared
                with the one obtained within the gradient approximation (dotted curve).}
\end{figure}

\begin{figure}[h]
\vspace{5mm}
\centerline{\includegraphics[width=13cm]{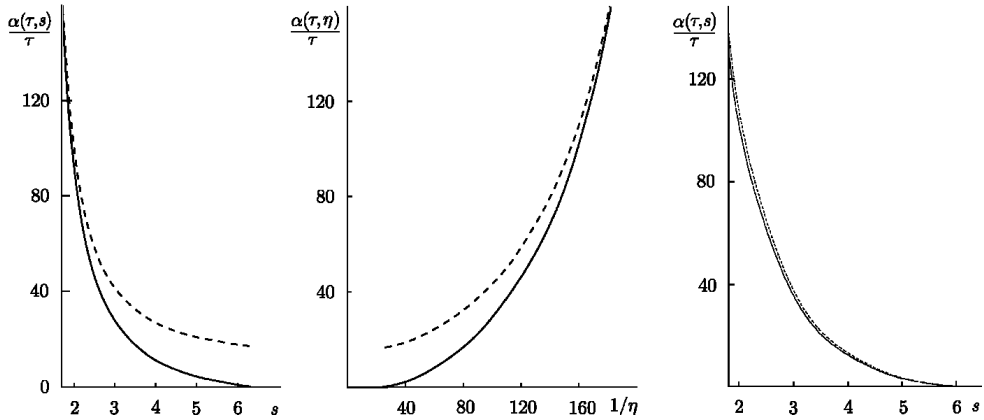}}
\caption{\small Gas--to--liquid nucleation barrier
                $\frac{\alpha(\tau,s)}{\tau}$ vs. $s$
                or
                vs. $\frac{1}{\eta}$
                at $\tau=0.6\tau_c({\cal {E}})$.
                Solid and dotted (gradient approximation) curves
                are obtained within the density functional approach.
                Dashed curves are obtained within the classical nucleation theory.}
\end{figure}

\newpage

\begin{figure}[h]
\vspace{5mm}
\centerline{\includegraphics[width=10cm]{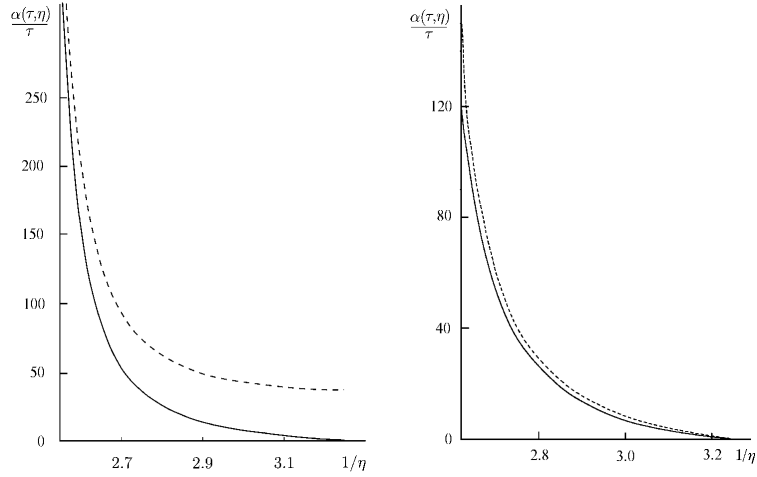}}
\caption{\small Liquid--to--gas nucleation (cavitation) barrier
                as a function of
                the density of the tensile liquid  at $\tau=0.6\tau_c({\cal{E}})$.
                Solid and dotted (gradient approximation) curves
                are obtained within the density functional approach.
                Dashed curve is obtained within the classical nucleation theory.}
\vspace{8mm}
\end{figure}

Let us discuss the obtained results.
The phase diagram of the considered fluid is shown in Figure~1.
The critical parameters are
\begin{eqnarray}
\label{006}
&&\rho_cv=\eta_c \approx 0.13044,
\nonumber\\
&&\tau_c({\cal {E}})\approx 0.00196518 \aleph^2(1+2\aleph^2{\cal{E}}^2),
\nonumber\\
&&\frac{p_c(\vert{\bf E}\vert)v}{E_1-E_0}
=\pi_c({\cal {E}})\approx 0.00009202 \aleph^2(1+2\aleph^2{\cal{E}}^2).
\end{eqnarray}
The electric field increases the critical temperature and pressure
as can be expected.
From the temperature dependence of the surface tension shown in the left panel in Figure~2
one concludes that it increases with switching on the field.
However, this change is conditioned only by the increase of the critical temperature
(\ref{006})
and the dependence
$\gamma$ vs. $\frac{T}{T_c(\vert{\bf{E}}\vert)}$
(as in the right panel in Figure~2)
is field independent.
The nucleation barriers $A$
calculated within the frames of the density functional theory
(exploiting the nonlocal grand thermodynamic potential density functional
or using the gradient approximation)
as well as within the classical nucleation theory
based on capillarity approximation \cite{001}
\begin{eqnarray}
\label{007}
\frac{A}{kT}=\frac{16\pi}{3}
\left(\frac{\gamma(T)}{kT}\right)^3
\frac{1}{\rho_l^2(T)\ln^2s}
\end{eqnarray}
(the supersaturation $s=\frac{p}{p(T)}$ characterizes the saturated vapour)
are shown in Figures~3,~4.
This quantity yields the nucleation rate
$J=J_0\exp\left(-\frac{A}{kT}\right)$
which is measured in nucleation experiments
in the cloud chamber
or in the upward thermal diffusion cloud chamber.
Similarly to the surface tension
the nucleation barriers appear to be field independent for given values of
$\frac{T}{T_c(\vert{\bf{E}}\vert)}$
and
$\frac{p}{p_c(\vert{\bf{E}}\vert)}$.

The results of a study of the nonuniform properties of atomic fluid in the electrical field
presented in this report
have been obtained within the frames of the standard density functional approach.
The new point
is the suggested density functional of the grand thermodynamic potential
(\ref{004}), (\ref{005}).
The main results concern the influence of electric field
on the phase diagram (Figure~1),
surface tension (Figure~2),
gas--to--liquid and liquid--to--gas nucleation barriers (Figures~3,~4).
These results may be useful
for analysis of the corresponding experimental data
(for example, for fluids of noble atoms)
on the nucleation phenomena in the presence of the electrical field.
However, it should be stressed
that the analysis of the influence of the electric field on the nucleation phenomena
in the concrete experiment
requires a separate consideration
that is out of the scope of the present paper.
The presented theory exhibits a kind of universality:
if the temperature and the pressure
are measured in units of their critical values
for a given strength of the electric field
the results for the surface tension and nucleation barriers are field independent.

\vspace{5mm}

A part of the paper was done in the Johannes Kepler Universit\"{a}t Linz.
V.~M.~Myhal thanks the \"{O}sterreich--Kooperation f\"{u}r Wissenschaft, Bildung und Kunst
for the fellowship
and H.~Iro for the hospitality.

\end{document}